\documentclass[10pt,pre,aps,twocolumn,showpacs]{revtex4}
\usepackage{graphicx}

\begin{document}

\title{Stochastic series expansion method \\ for quantum Ising models with 
arbitrary interactions}

\author{Anders W. Sandvik} 
\affiliation{Department of Physics, {\AA}bo Akademi University, 
Porthansgatan 3, FIN-20500, Turku, Finland}

\date{\today}
\begin{abstract}
A quantum Monte Carlo algorithm for the transverse Ising model with arbitrary
short- or long-range interactions is presented. The algorithm is based on 
sampling the diagonal matrix elements of the power series expansion of the 
density matrix (stochastic series expansion), and avoids the interaction 
summations necessary in conventional methods. In the case of long-range
interactions, the scaling of the computation time with the system size $N$ 
is therefore reduced from $N^2$ to $N \ln{(N)}$. The method is tested on a 
one-dimensional ferromagnet in a transverse field, with interactions 
decaying as $1/r^2$.
\end{abstract}


\maketitle

\section{introduction}

Monte Carlo studies of classical and quantum many-body systems with long-range 
interactions are limited by time-consuming summations over the interacting 
particle pairs, the number of which grows quadratically with the system size. 
Many important problems in both basic and applied science can be mapped onto
long-range interacting spin models, and hence it would be desirable to develop
more efficient numerical techniques for tackling them. For classical Ising
models, considerable progress has indeed been made on algorithms scaling
almost linearly with the system size \cite{luijten}. In the context of 
simulated annealing \cite{annealing}, where the ground state of a classical 
system (typically with complicated interactions) is obtained through a 
simulation where the temperature is slowly lowered to zero, it has been
suggested \cite{qann} that a more rapid convergence could be achieved 
by using a quantum model, e.g., the Ising model in a transverse (spin 
flipping) field. Even in an imaginary-time path-integral formulation,
the quantum fluctuations can, at least in some cases \cite{santoro}, 
relax the system towards its classical ground state more rapidly than 
thermal fluctuations. This is a strong motivation for developing 
more efficient simulation methods for quantum Ising models. Another important 
reason is the continued prominence of the transverse Ising model in the theory
of magnetism, particularly in the context of quantum phase transitions 
\cite{sachdev,guo,rieger}. 
Whereas transverse Ising models with short-range 
interactions have recently been actively studied using quantum Monte Carlo 
methods \cite{guo,rieger}, numerical work on long-range models has so 
far been limited to special cases \cite{rozenberg}. In some of the best 
experimental realizations of the transverse Ising model the interactions
are in fact long-ranged \cite{lihof}.

Here a stochastic series expansion (SSE) \cite{sse} algorithm for transverse
Ising models with long-range interactions is introduced in which the direct 
summation over the interacting spins is avoided. The computation time scales 
with the system size N as $N \ln{(N)}$ times the spatial integral of the 
absolute value of the interaction [which normally converges as $N \to \infty$,
or diverges only as $\ln{(N)}$]. Both local and cluster-type updates 
are developed for the transverse Ising model with 
arbitrary interactions. The cluster update is a generalization of the 
classical Swendsen-Wang cluster method \cite{swendsen} to the transverse 
Ising model, and shares some features with a scheme previously used within the 
continuous-time world-line algorithm \cite{rieger}. The way to treat the 
long-range interactions generalizes the scheme developed for the classical 
Ising model by Luijten and Bl\"ote \cite{luijten,rnote}. The integration of 
these features in the SSE formalism should open new opportunities for detailed 
numerical studies of a wide range of important models. The algorithm is 
here tested on a ferromagnetic chain with interactions decaying as $1/r^2$, 
for which results in the classical limit are available for comparison 
\cite{anderson,kosterlitz,glumac,cannas,luijten2}.

In Sec.~II the application of the SSE method to the transverse Ising
model is described in detail. Local updates as well as classical and
quantum-cluster updates are discussed. Results for the model with $1/r^2$ 
interactions are presented in Sec.~III. Sec.~IV concludes with a brief 
discussion.

\section{STOCHASTIC SERIES EXPANSION}

The SSE method \cite{sse} is an efficient alternative to worldline quantum 
Monte Carlo \cite{worldline}. It is based on a generalization of the 
power-series scheme for the Heisenberg ferromagnet that was developed by 
Handscomb in the early 1960's \cite{handscomb}. Handscomb's method was later 
extended to some other models \cite{hgen}, but the requirement of 
analytically calculable traces of the terms of the 
expansion inhibited further progress. In the SSE 
method, a basis is instead chosen, and the traces are also evaluated 
stochastically, in combination with the sampling of the operator products 
in the series expansion of $\rm{exp}(-\beta H)$. This starting point
for quantum Monte Carlo is as generally applicable as the 
worldline (imaginary-time path-integral) approach. 
Recently, loop-type cluster updates \cite{evertz} have been developed and 
generalized for efficient SSE simulations of a wide range of models 
\cite{sseloop1,sseloop2}. However, since the loop updates rely heavily on the 
presence of off-diagonal pair (or multi-particle) interactions, they 
cannot be directly adapted to the transverse Ising model in the standard basis
where the Ising term is diagonal. In the basis where the field is diagonal, 
loop updates can be easily implemented \cite{sseloop1,sseloop2}
but then sign problems \cite{ssesign} appear when the interaction is 
frustrated. Here the SSE method is applied to an arbitrary transverse Ising 
model, i.e., with no limitations on the sign and range of the spin-spin 
interaction. Several types of local and cluster-type updates will be 
described.

\subsection{Configuration space}

Consider the general Hamiltonian for the Ising model in a transverse field
of strength $h$,
\begin{equation}
H = \sum\limits_{i,j} J_{ij}\sigma^z_i\sigma^z_j - h\sum\limits_i \sigma^x_i ,
\label{hamiltonian}
\end{equation}
where ${\bf \sigma}_i$ is a Pauli spin matrix ($\sigma^z_i = \pm 1$) and 
$J_{ij}$ is the strength of the interaction between spins $i$ and $j$, which
can be random or uniform and of any sign. The dimensionality is arbitrary.
Define the operators
\begin{eqnarray}
H_{0,0} & = & 1, \label{ha} \\
H_{i,0} & = & h(\sigma^+_i + \sigma^-_i),~~~~ i > 0, \label {hc} \\
H_{i,i} & = & h, ~~~~ i > 0, \label {hb} \\
H_{i,j} & = & |J_{ij}| - J_{ij}\sigma^z_i\sigma^z_j, ~~~~ i,j > 0,~i \not= j. 
              \label{hd} 
\end{eqnarray}
Up to a constant, the Hamiltonian can be written as
\begin{equation}
H = -\sum\limits_{i=1}^N \sum\limits_{j=0}^N H_{i,j} .
\label{hsum}
\end{equation}
The constants $H_{i,i}$ are introduced for purposes that will become
clear below. Note that $H_{0,0}$ is not included as a term in the 
Hamiltonian (\ref{hsum}) but will be important in the simulation scheme. 

In the SSE approach \cite{sse} to finite-temperature quantum Monte Carlo, the 
partition function $Z={\rm Tr}\lbrace {\rm exp}(-\beta H)\rbrace$ is written 
as a power-series expansion, with the trace expressed as a sum over diagonal 
matrix elements in a suitably chosen basis. Using (\ref{hsum}) then gives
\begin{equation}
Z = \sum\limits_{\alpha} \sum\limits_{n=0}^\infty \sum\limits_{S_n}
{\beta ^n \over n!} 
\left \langle \alpha \right | \prod\limits_{l=1}^n H_{i(l),j(l)}
\left | \alpha \right \rangle ,
\label{zn}
\end{equation}
where $S_n$ denotes a sequence of $n$ operator-index pairs (hereafter
referred to as operators):
\begin{equation}
S_n = [i(1),j(1)],\ldots ,[i(n),j(n)],
\end{equation}
with $i(l) \in \lbrace 1,\ldots,N\rbrace$ and 
$j(l) \in \lbrace 0,\ldots,N\rbrace$. The standard basis 
$\lbrace |\alpha \rangle \rbrace = \lbrace |\sigma^z_1 ,\ldots ,
\sigma^z_N \rangle \rbrace$ is used.

Because of the constants added to $H_{i,j}$ in (\ref{hd}), the eigenvalues of
these operators are $2|J_{ij}|$ and $0$. All non-zero terms in (\ref{zn}) are 
therefore positive and can be used as relative probabilities in an importance 
sampling scheme. A term is specified by a state $|\alpha \rangle$ and an 
operator sequence $S_n$. One can show that the total internal energy 
(including the constants added to $H$) is given by \cite{sse,handscomb} 
$E=-\langle n\rangle /\beta$. Hence, the size of the operator sequence to 
be stored in computer memory scales as $\beta N I_N(J)$, where 
\begin{equation}
I_N(J)={1\over N}\sum\limits_{i=1}^N\sum\limits_{j=1}^N |J_{ij}| ,
\label{in}
\end{equation}
which converges or grows much slower than $N$ for most cases of interest.

In order to construct an efficient sampling scheme, it is useful to cut the 
expansion (\ref{zn}) at some power $n=L$, sufficiently high for the remaining
truncation error to be exponentially small and completely negligible [$L$ 
clearly has to be $\sim \beta N I_N(J)$]. One can then obtain an expansion
for which the length of the operator sequence is constant, by considering 
random insertions of $L-n$ unit operators $H_{0,0}$ in the product 
in (\ref{zn}). Adjusting for the $L \choose n$ possible insertions gives
\begin{equation}
Z = {1\over L!}\sum\limits_{\alpha} \sum\limits_{S_L}
\beta ^n (L-n)!
\left \langle \alpha \right | \prod\limits_{l=1}^L H_{i(l),j(l)}
\left | \alpha \right \rangle ,
\label{zl}
\end{equation}
where $[i(l),j(l)] = [0,0]$ is now also an allowed operator in the
sequence $S_L$, and $n$ denotes the number of non-$[0,0]$ operators. Note 
again that $H_{0,0}$ is not part of the Hamiltonian, but is introduced 
only for the purpose of constructing a computationally simpler updating 
scheme where the operator list has a fixed length. 

It is useful to define states 
$|\alpha (p) \rangle = |\sigma^z_1(p),\ldots,\sigma^z_N(p)\rangle$ obtained 
by propagating $|\alpha \rangle = | \alpha (0)\rangle$ by the first $p$ 
operators in $S_L$;
\begin{equation}
|\alpha (p) \rangle = r\prod\limits_{l=1}^p H_{i(l),j(l)} |\alpha \rangle ,
\label{propagated}
\end{equation}
where $r$ is a normalization factor. A non-vanishing matrix element in 
(\ref{zl}) then corresponds to the periodicity condition $| \alpha (L) 
\rangle = |\alpha (0) \rangle$, which requires that for each site $i$ there is 
an even number (or zero) of spin flipping operators $[i,0]$ in $S_L$. 
Definition (\ref{hd}) implies that the Ising operators $[i,j]$ may act only 
on states with $\sigma^z_i = \sigma^z_j$ if $J_{ij} < 0$ (ferromagnetic), 
or $\sigma^z_i = -\sigma^z_j$ if $J_{ij} > 0$ (antiferromagnetic). There 
are no other constraints. 

\begin{figure}
\includegraphics[width=4.5cm]{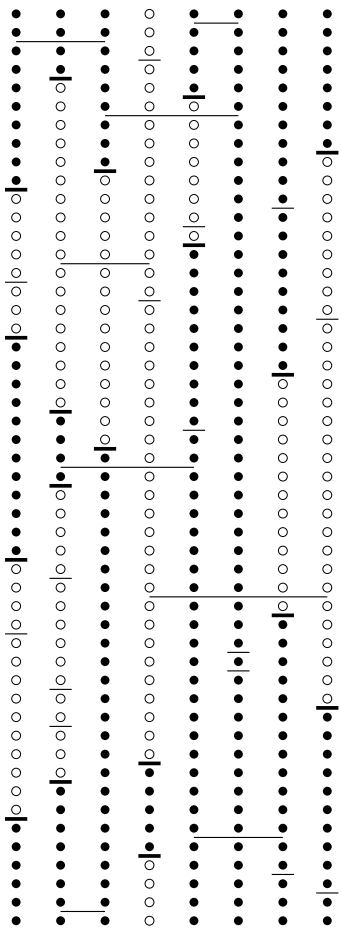}
\caption{An SSE configuration for an 8-site one-dimensional system. Here
the truncation $L=49$, and the expansion order of the term (i.e., the number
of Hamiltonian operators present) $n=40$. The solid and open circles represent 
the spins $\sigma^z_i(p)=\pm 1$, with the propagation index $p=0,\ldots,L$ 
corresponding to the different $8$-spin rows. The thick and thin short 
horizontal bars represent spin-flip operators $H_{i,0}$ and constants 
$H_{i,i}$, respectively. The longer lines represent Ising operators 
$H_{i,j}$ ($i\not =j$) acting on the spins at the line-ends.}
\label{conf}
\end{figure}

An SSE configuration is illustrated in Fig.~\ref{conf}. The vertical
direction in this representation will be referred to as the SSE {\it 
propagation direction}. It can be related to the imaginary-time direction 
in standard path integral representations \cite{irsse}. Note that this full 
configuration, including all the states $|\alpha (p) \rangle$ explicitly, 
does not have to be stored in the simulation. A single state and the operator
sequence suffice for reproducing all the states, and such a representation 
is used in some stages of the simulation. For some updates it is convenient 
to generate other representations, as will be discussed below.

\subsection{Local updates}

The sampling of Eq.~(\ref{zl}) can be carried out using simple operator 
substitutions of the types
\begin{eqnarray}
&& [0,0]_p \longleftrightarrow [i,j]_p, ~~~ i,j \not=0,
\label{upd1} \\
&& [i,i]_{p_1}[i,i]_{p_2} \longleftrightarrow 
[i,0]_{p_1}[i,0]_{p_2}, ~~~ i \not=0,
\label{upd2}
\end{eqnarray}
where the subscript $p$ indicates the position ($p=1,\ldots ,L$) of the 
operator in the sequence $S_L$. The power $n$ is changed by $\pm 1$ in the 
{\it diagonal update} (\ref{upd1}) and is unchanged in the {\it off-diagonal 
update} (\ref{upd2}). In the diagonal update the Ising terms $[i,j]$ and the 
constants $[i,i]$ are sampled. The constants are used in the off-diagonal
update as a means of achieving easy insertions and removals of 
two spin-flipping operators $[i,0]$. With the value $h$ chosen for the 
constant in (\ref{hb}), the operator replacements do not change the 
weight of the SSE configuration. However, the off-diagonal update also 
leads to spin flips in the propagated states between $p_1$ and $p_2$; 
$\sigma^z_i (p_1),\ldots,\sigma^z_i (p_2-1)$ $\to$
$-\sigma^z_i (p_1),\ldots,-\sigma^z_i (p_2-1)$.
[$p_1 > p_2$ also has to be considered, leading to 
flipped $\sigma^z_i (p_1),\ldots, \sigma^z_i (L-1)$ $\sigma^z_i (0),\ldots,
\sigma^z_i (p_2-1)$], which is allowed if (and only if) no Ising operators 
acting on site $i$ are present in $S_L$ between positions $p_1$ and $p_2$. 
{\it Note that this constraint is completely local, regardless of the range 
of the interaction, and that the update requires no knowledge of the spin 
state}. This is the reason for the advantage of this simulation scheme over 
worldline methods \cite{worldline,rieger}, where calculating the acceptance 
probability for every update requires a summation over all the spins 
interacting with those flipped. Here an allowed off-diagonal update 
(\ref{upd2}) leaves the weight unchanged and can be carried out with 
probability one. 

If $h \not=0$, the above updates of the operator sequence suffice for 
achieving ergodicity. If there are no Ising operators acting on a site $i$, 
$\sigma^z_i(0),\ldots,\sigma^z_i(L-1)$ can also be flipped without changes in 
$S_L$. This update in principle makes simulations using the present scheme 
possible also for $h=0$, but in practice unconstrained spins occur frequently
only at high temperatures, when $\langle n\rangle$ is small. Other types of 
``classical'' spin flips --- flips of clusters --- are also possible, and 
will be discussed in Sec.~II C.

The simulation can be started with a random state $|\alpha (0)\rangle$ and 
a sequence $S_L$ containing only $[0,0]$ operators. The truncation $L$ can be 
chosen arbitrarily (small); it is adjusted during the equilibration part of 
the simulation, e.g., by requiring $L > (4/3)n$ after each update. This 
ensures than $n$ never reaches $L$ during the reminder of the simulation,
and hence that there will be no detectable systematic errors arising from 
the truncation of the expansion \cite{sse}. In the beginning of an 
updating cycle, the operator sequence $S_L$ and the state $|\alpha (0) 
\rangle$ is stored. 

The diagonal update (\ref{upd1}) is 
attempted successively for all $p = 1,\ldots ,L$. In the course of this 
process, the spin state is propagated by flipping spins $\sigma^z_i$ as 
off-diagonal operators $[i,0]$ are encountered in $S_L$, so that the states 
$|\alpha (p) \rangle$ are generated successively. For an $[i,j] \rightarrow 
[0,0]$ update, i.e., removing a Hamiltonian operator, there are no constraints
and the update should always be accepted with some non-zero probability. In 
the case of $[0,0] \rightarrow [i,j]$, i.e., inserting an operator from the 
Hamiltonian, there are constraints, and the update may not be allowed for 
all $i,j$. However, initially the indices $i,j$ are left undetermined and 
it is assumed that any $[i,j]$ would be allowed. Under this assumption, the 
acceptance probabilities for the diagonal update are given by
\begin{eqnarray}
P([0,0] \to [i,j]) & = & {\beta (Nh+2\sum_{ij}|J_{ij}|)\over L-n}, \\
P([i,j] \to [0,0]) & = & {L-n+1\over \beta (Nh+2\sum_{ij}|J_{ij}|)},
\end{eqnarray}
where $\sum_{ij}$ does not include $i=j$ and $P>1$ should be interpreted as 
probability one, as usual. These probabilities are simply obtained from the 
ratio of the new and old prefactor in (\ref{zl}) when $n \to n \pm 1$;
\begin{equation}
\beta^{\pm 1}{[L-(n \pm 1)]! \over (L-n)!},
\end{equation}
and the ratio between the matrix element $1$ of the $[0,0]$ operator and 
the sum $Nh+2\sum_{ij}|J_{ij}|$ of the non-zero matrix elements of all 
$[i,j]$ operators. Staying with the assumption that any $[i,j]$ is allowed 
in the update $[0,0] \rightarrow [i,j]$, the 
relative probability of an operator with the first index $i$ is 
$P(i)=\sum_j M_{ij}$, where $M_{ij}$ is the non-zero matrix element 
corresponding to $H_{ij}$ (i.e., $h$ for $i=j$ and and $2|J_{ij}|$ else). 
The normalized cumulative probabilities $P_c(k=1,\ldots ,N)$ are stored 
in a pre-generated table;
\begin{equation}
P_c(k)={\sum_{i=1}^k P(i) \over \sum_{i=1}^N P(i)}.
\end{equation}
In order to select the first index $i$ of the operator $[i,j]$ to be 
inserted, a random number $0 \le R < 1$ is generated. The table $P_c$ 
is searched (using, e.g., a simple binary search) for 
the smallest $k$ for which $P(k) \ge R$; the first index
of the operator $[i,j]$ is then $i=k$. The second index can be chosen
in a completely analogous way, with the relative probability for $j$, given
$i$, being $M_{ij}$. For a random system with long-range interactions,
a pregenerated table with $N^2$ elements is hence needed for storing all
the cumulative probabilities for the second index. For non-random interactions
in a translationally invariant system, the first index can be selected at 
random with equal probabilities without searching a table, and the size of
the second table is reduced to $N$. For a short-range or truncated interaction
the table size is smaller, corresponding to the number of spins within the
range of the interaction; clearly, the whole selection process should then
be reduced to a single step for obtaining both $i$ and $j$ (e.g., selecting
one out of a total number $\sim N$ of operators and reading the corresponding
$i,j$ from a table). The two-step procedure is advantageous for non-random 
long-range interactions, where it allows for the reduction of the size of
the probability table from $N^2$ to $N$. For random models, the storage
requirement is always $N^2$, and it may then again be better to combine the 
first and second index searches, using a single size-$N^2$ table for all 
the cumulative probabilities of $[i,j]$. For short-range random interactions
the size of the table is $N$ times the number of spins within the 
interaction range.

The operator $[i,j]$ generated as above may or may not be allowed in the 
current spin configuration $|\alpha (p) \rangle$. If $\sigma^z_i(p)$ and 
$\sigma^z_{j} (p)$ indeed are in an allowed state, $[i,j]$ is inserted at 
position $p$. Otherwise, the process for generating $[i,j]$ has to be 
repeated, until an allowed operator has been generated. The reject-and-repeat
step leads to the correct probabilities for selecting among all the allowed 
diagonal operators $[i,j]$. Typically, an allowed operator is generated very 
quickly, since the interactions favor the allowed spin alignment. Note that 
the constants $[i,i]$ are always allowed (for $h >0$), so there is no risk 
of the search never terminating. 

The off-diagonal update (\ref{upd2}) can be efficiently carried out if 
$S_L$ is first partitioned into separate subsequences for each site $i$. 
Subsequence $i$ contains only spin-flipping operators $[i,0]$ and constants 
$[i,i]$. Their positions in $S_L$ are also stored, to be used for recombining
the subsequences after the update. The constraints on modifications at site 
$i$ imposed by Ising operators $[i,j]$ or $[j,i]$ (for any $j$) can be stored
as flags indicating the presence of one or several of these operators between 
neighboring subsequence operators. Updating a subsequence amounts to 
selecting two non-constrained neighboring operators at random from the 
subsequence, and carrying out the substitution (\ref{upd2}) if the two 
operators are identical. If they are different, they can be permuted. A 
number proportional to the subsequence length of such pair updates are 
carried out for each subsequence, after which they are recombined 
into a new $S_L$. 

The diagonal update (\ref{upd1}) at all positions in $S_L$ require 
$\sim L \ln{(N)} \sim \beta N\ln{(N)}I_N(J)$ operations, where the factor 
$\ln{(N)}$ is the scaling of the average number of operations needed to 
search the cumulative probability table(s) in the case of long-range
interactions. Partitioning $S_L$ into subsequences and updating all of them 
according to (\ref{upd2}) requires on the order of $L$ operations. Hence, 
the number of operations for a full updating cycle of the degrees of freedom 
of the system (one Monte Carlo step) scales as $\beta N\ln{(N)}I_N(J)$. This 
should be compared to the $\beta N^2$ scaling in worldline methods 
\cite{worldline,rieger}, where one power of $N$ is due to the summation 
required  to calculate the weight change when flipping a spin interacting 
with $N$ other spins. Here this summation has been circumvented by writing 
the interactions in the SSE formalism as fluctuating constraints that are 
purely local. 

\subsection{Classical cluster update}

In the Swendsen-Wang cluster algorithm \cite{swendsen} for the classical Ising
model, i.e., with $h=0$ and a uniform nearest-neighbor interaction of strength
$J$, auxiliary bond variables $b_{ij}$ are introduced in order to construct 
clusters of spins that can be flipped independently of each other. Given a 
spin configuration, and with initially all bond variables $b_{ij}=0$, for 
every interacting spin pair for which $\sigma_i \sigma_j = -J/|J|$ 
(i.e., the orientation energetically favored) the bond variable is set, 
$b_{ij}=1$, with probability $P=1-{\rm e}^{-2|J|\beta}$. When all bonds 
have been visited, clusters of spins connected by $b_{ij}=1$ bonds are formed,
and each of these clusters is flipped with probability $1/2$. Single 
spins not connected to any $b_{ij}=1$ bond are single-spin clusters. After 
the clusters have been flipped, all the bond variables are again set to zero 
and the process is repeated. This scheme can in fact be constructed using 
the SSE formalism, as an alternative to the Fortuin-Kastelin mapping 
\cite{ft}, on which the Swendsen-Wang algorithm is based.

The relation to the Swendsen-Wang algorithm is shown as follows, by applying
the SSE method to the classical Ising model, now again considering a general 
form of the interaction $J_{ij}$ and with the bond-operator $H_{ij}= |J_{ij}|
- J_{ij}\sigma^z_i\sigma^z_j$ as in Eq.~(\ref{hd}). Since all operators  
$H_{ij}$ commute, the operator ${\rm e}^{-\beta H}$ can be written as a 
product of operators ${\rm e}^{\beta H_{ij}}=1 + \beta H_{ij} + \ldots$. The 
uniqueness of the power-series expansion then implies that in the SSE, where 
${\rm e}^{-\beta H}$ is expanded directly, the probability of having one or 
more operators $H_{ij}$ on a bond $i,j$ when $\sigma_i \sigma_j = 
-J_{ij}/|J_{ij}|$ is $1-{\rm e}^{-2|J_{ij}|\beta}$, i.e., exactly the 
probability of having the bond variable $b_{ij}=1$ in the Swendsen-Wang 
scheme. In a configuration $\sigma_i \sigma_j = J_{ij}/|J_{ij}|$ there can 
be no operators on the bond in the SSE, and the Swendsen-Wang $b_{ij}=1$
probability is also zero per construction. One can hence make the connection 
that one or more operators acting on a spin pair in the SSE scheme corresponds 
to a filled bond ($b_{ij}=1$) in the Swendsen-Wang algorithm. The definition 
of a cluster is then exactly the same in the two algorithms. Clearly, such
a cluster in the SSE can also always be flipped, since the Ising operators 
only impose constraints on the relative orientations of connected spins, 
which is maintained when the cluster is flipped. 
Since the weight does not change, the 
flip should be done with probability $1/2$. The scheme is hence identical 
to the Swendsen-Wang algorithm, except that the filled bonds  $b_{ij}=1$ 
in SSE are generated in a different way, using the diagonal update 
(\ref{upd1}). Note that for a classical model, all the propagated SSE states 
(\ref{propagated}) are identical, i.e., $\sigma^z_i(p) = \sigma^z_i(0)$ for 
all $p=0,\ldots,L-1$, and hence no state propagations have to be considered
as the diagonal update is carried out. 

It is interesting to note that the SSE scheme for the classical Ising model
should in fact be more efficient than the standard Swendsen-Wang algorithm
at high temperatures. This is because the number of operators in the SSE
operator list scales as $E(T)/T$, where $E(T)$ is the total energy at 
temperature $T$ ($E \sim N$) and for large $T$ the construction of the 
clusters based on the operator list should then be faster than visiting
all the bonds, as is done in the Swendesn-Wang algorithm. 
However, in practice the 
interesting physics occurs when the number of SSE operators per interacting
spin pair is of the order of one or larger, and then there are no advantages
of the SSE classical cluster algorithm relative to Swendesen-Wang.

The classical SSE cluster update can also be used in the presence of the 
transverse field ($h > 0$). The clusters are defined in terms of bonds 
signifying the presence of one or more Ising operator, as above, without 
regard for the single-spin flipping operators $H_{i,0}$ and constants
$H_{i,i}$. These operators can be neglected because when a cluster is flipped,
all  spins $\sigma^z_i$ belonging to the cluster are implicitly flipped in all 
propagated states (\ref{propagated}), i.e., $\sigma^z_i(p)$ $\to$ 
$-\sigma^z_i(p)$ for all $p=0,\ldots,L-1$ (this is the reason for the
term ``classical cluster'' even when $h > 0$) and hence all operations with 
the single-spin operators remain valid and produce the same factors in the 
weight before and after the cluster flips. Note again that only the first 
state, i.e., $\sigma^z_i(0)$, $i=1,\ldots,N$, has to be stored when 
constructing the classical clusters.

\begin{figure}
\includegraphics[width=5cm]{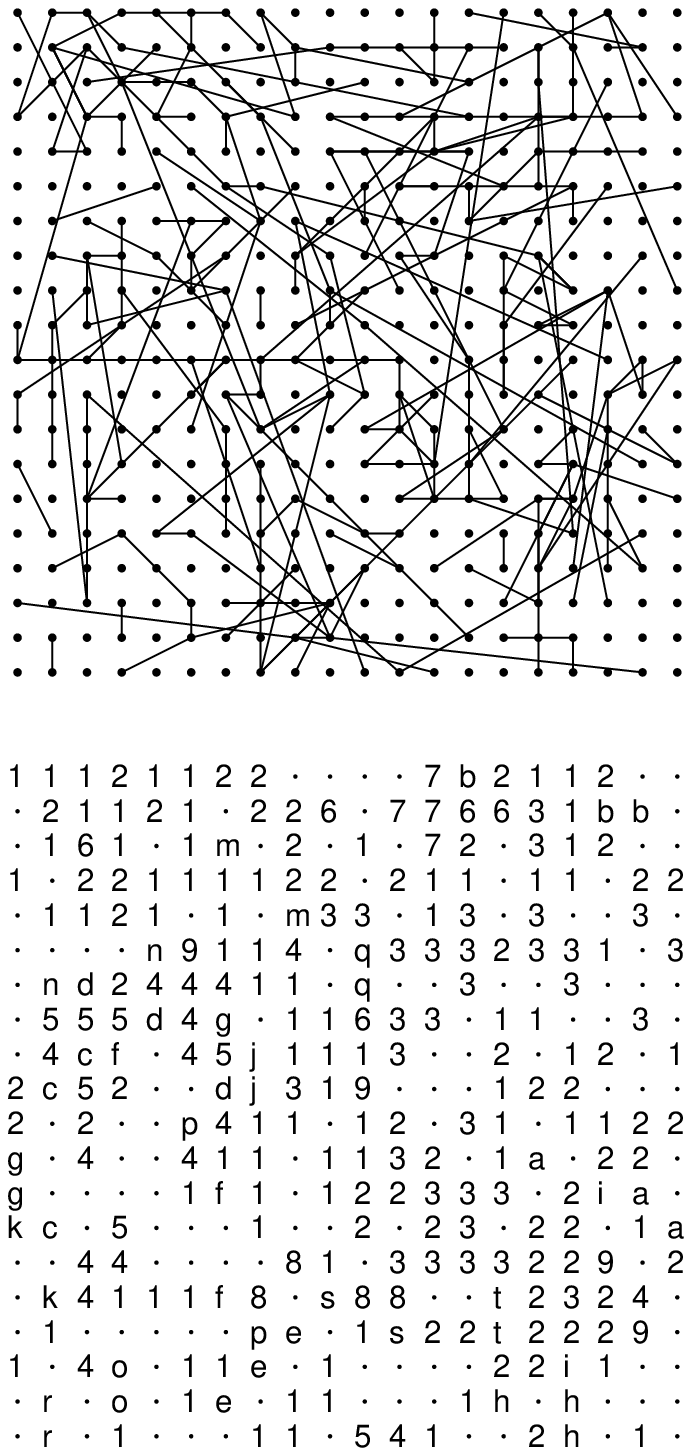}
\caption{Upper panel: Interaction bonds in a configuration for a 2D system
with long-range interactions. Lower panel: The clusters constructed from 
the bonds. Sites with equal symbols belong to the same cluster. 
Dots indicate spins not acted on by any Ising operator and 
constitute single-spin clusters.}
\label{c_cluster}
\end{figure}

In the case of long-range interactions, a cluster can consist of several 
intertwined pieces on the lattice, as illustrated for a two-dimensional case
in Fig.~\ref{c_cluster}. Regardless of the range of the interaction, 
the construction of the clusters,
given an SSE operator list, can be easily carried out using a number of 
operations scaling as the number of operators in the list.

Since the classical SSE cluster update is equivalent to the Swendsen-Wang 
algorithm in the classical limit and only takes the Ising terms into account
also in the quantum case, it cannot be expected to be efficient much beyond 
the classical limit $h=0$. For a non-random system that undergoes a phase
transition at $T_c(0)$ when $h=0$, the critical temperature is reduced by
the transverse field; $T_c(h) < T_c(0)$. Hence, the classical clusters will
percolate for $T > T_c$ and this update will not be efficient close to $T_c$.
The primary reason to introduce the classical cluster update here was to
demonstrate the relationship between SSE and the Swendsen-Wang algorithm.
In the case of long-range interactions, the scheme becomes very similar
to the Luijten-Bl\"ote algorithm \cite{luijten}, again just differing in 
the way the bonds are generated.

\subsection{Quantum-cluster update}

The purpose of the quantum-cluster update is to effect flips of spins 
$\sigma^z_i(p)$ only in a limited number of propagated states $p$, in
different states for different sites $i$. In other words, these clusters
will be finite and irregularly-shaped both in the space and SSE propagation 
(imaginary time) direction. In the process, operator substitutions $H_{i,i}
\leftrightarrow H_{i,0}$ (constant to spin-flip, and vice versa) will also
be accomplished. This update hence replaces the local off-diagonal update
(\ref{upd2}).

\begin{figure}
\includegraphics[width=4.7cm]{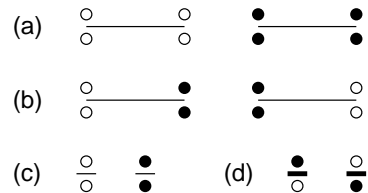}
\caption{All the possible 4-leg and 2-leg vertices. (a) Ferromagnetic
Ising vertices, (b) antiferromagnetic Ising vertices, (d) constant
vertices, and (d) spin-flip vertices.}
\label{vertices}
\end{figure}

To discuss the quantum-cluster update, 
it is useful to introduce the notion of {\it vertices}
\cite{sseloop1,sseloop2}. 
Looking at the graphical representation of a configuration 
in Fig.~\ref{conf}, it can be noted that the vertical ``lines'' of same spins 
between two operators acting on a given site constitute redundant information.
The full configuration can be represented by a list of positions (on the 
lattice) of the operators, and the spin states (on one or two sites for the 
model considered here) before and after the operators act. These relevant 
spins are called {\it legs} of the 2-spin vertices (corresponding to constant 
and spin-flip operators) or 4-spin vertices (corresponding to Ising 
bond-operators). All possible vertices for the transverse Ising model are 
shown in Fig.~\ref{vertices}. Note that only those Ising vertices that are 
compatible with the sign of the interaction between a given pair of spins 
are allowed for those spins; again, this is due to the choice of constant 
in the bond-operator (\ref{hd}). In the computer, the vertices are linked 
to each other by pointers, so that from a given vertex-leg one can reach the 
next or previous vertex that has a leg on the same site (i.e., there are 
links that replace the segments of vertical lines of same spins in 
Fig.~\ref{conf}). A detailed discussion of the practical implementation of 
a linked vertex list has been given in Ref.~\onlinecite{sseloop2}.

\begin{figure}
\includegraphics[width=6.2cm]{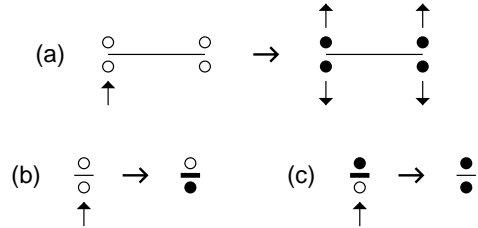}
\caption{Examples of vertex processes: (a) reversal of a ferromagnetic Ising
vertex, (b) constant to spin-flip, and (c) spin-flip to constant.}
\label{q_cluster}
\end{figure}

To construct and flip a quantum-cluster, one of the legs of one of 
the $n$ vertices 
is picked at random, and the corresponding spin is flipped. Depending on the 
type of the vertex, different actions are taken, examples of which are given 
in Fig.~\ref{q_cluster}. The arrow pointing into the vertex indicates the 
{\it entrance leg}. In the case of an Ising vertex, all the four spins are 
flipped and the cluster building process branches out from all the legs, 
as indicated by the arrows pointing out from the vertex. Using the pointers 
of the linked vertex list, the arrows point to legs of other vertices; these 
become new entrance legs which are put on a stack and subsequently processed 
one-by one. If the entrance leg is on a constant or spin-flip vertex, 
only the entrance spin is flipped. The 
vertex type then also changes, in terms of operators from $H_{i,0}$ to 
$H_{i,i}$, and vice versa. In these cases there is no branching-out and no 
new legs are put on the stack, i.e., this particular branch of the cluster 
terminates. If a link points to a spin that has already been flipped (i.e.,
two arrows point toward each other), that leg should not be used 
again as an entrance and is hence not put on 
the stack. Therefore, each vertex-leg can be visited at most once (each spin 
can be flipped at most once) and the cluster is completed when there are no 
more entrance-legs on the stack. The reason that the cluster can always be 
flipped is again that the SSE weight is not affected; the matrix element of 
the Ising bond-operator is not affected when both spins are flipped (in the 
absence of an external field in the $z$-direction, which would necessitate 
a modified approach), and the matrix elements for the constant and spin-flip 
operators are both equal to $h$.

The construction of a single cluster, which is flipped with probability one,
is a quantum-mechanical analogue of the classical Wolff algorithm 
\cite{wolff}; in the absence of the transverse field the clusters are 
identical to those of the Wolff algorithm. Note, however, that there is a 
difference when constructing more than one cluster: The number of operators
in the SSE operator list, and their positions on the lattice, do not change 
in the quantum-cluster update. The clusters are therefore completely 
deterministic once the operator list is given. Hence, when constructing 
several clusters using the same SSE operator list, 
it is quite likely that the same cluster is constructed and flipped multiple 
times. This is clearly not efficient. However, one can also construct all 
clusters, as in the Swendsen-Wang scheme, and only flip them with probability
$1/2$. This is done by always starting a new cluster from a vertex-leg which 
has not yet been visited. Every vertex-leg belongs uniquely to one cluster, 
and clearly the number of operations required to complete this updates 
then scales as $L$, i.e., typically as $\beta N$.

A natural definition of a Monte Carlo step including the quantum-cluster 
update is a full sweep of diagonal updates, followed by the construction of 
the linked list of vertices, in which  all clusters are constructed and flipped
with probability $1/2$. After that, the updated vertex list is mapped back 
into a state $|\alpha (0)\rangle$ and an operator sequence $S_L$. Free 
spins, i.e., those that are not acted on by any operators, can again be 
considered as single-spin clusters and should also be flipped with probability
$1/2$. No local off-diagonal updates (\ref{upd2}) are needed. 

Since the quantum-cluster update explicitly includes the quantum mechanical
features of the configurations (i.e., the presence of spin-flip operators), 
it can be expected to work well also close to a quantum phase transition 
($T_c=0$) driven by varying $h$.

\section{1D INVERSE-SQUARE FERROMAGNET}

As a non-trivial demonstration of the method, a ferromagnetic chain with 
interactions decaying as $1/r^2$ is considered next. The interaction is summed
over all $i,j$ in (\ref{hamiltonian}), i.e., each pair is counted twice. 
Periodic boundary conditions are used. $J_{ij}$ includes both distances
in the periodic system, i.e., 
\begin{equation}
J_{ij} = J_{ji} = {J\over 2} \left 
({1\over |i-j|^{2}} + {1 \over (N-|i-j|)^{2}} \right ),
\end{equation}
where $J$ sets the over-all energy scale. 

The classical $1/r^2$ Ising chain has been the subject 
of numerous studies \cite{anderson,kosterlitz,glumac,cannas,luijten2}. 
The long-range interaction allows for a finite-$T$ phase transition even in 
one dimension. The transition is of an unusual kind, with the correlation 
length exponent $\nu = \infty$, and a discontinuous jump in the magnetization
at $T_c$. It can be thought of as a one-dimensional analogue of the 
Kosterlitz-Thouless transition, with the topological excitations being kink 
solitons \cite{kosterlitz}. The model is also important because it can be 
mapped onto the Kondo problem \cite{anderson}.

For small $h/J$, one can expect a behavior similar to the classical case,
i.e., a finite-$T$ phase transition to a ferromagnetic state. For $h \to
\infty$ the system becomes disordered, and there should therefore be a 
finite $h_{\rm c}$ for which the system undergoes a quantum phase transition
(i.e., $T_{\rm c}=0$). For $h < h_c$, $T_c > 0$ and one can expect the same 
universality class as in the classical case, since the quantum fluctuations 
become irrelevant at $T_{\rm c}$. Here only a single field-strength 
$h/J = 0.5$ is considered. The simulations  show that $T_{\rm c} > 0$ in 
this case.

The model is invariant with respect to flipping all spins, which means
that for any finite system the average magnetization vanishes. The squared
magnetization,
\begin{equation}
M^2 = \left \langle \left ( 
{1\over N} \sum_{i} \sigma^z_i
\right )^2 \right \rangle ,
\end{equation}
is therefore calculated. Results for $M^2$ with statistical errors in the 
fifth decimal place can easily be obtained for systems with several hundred 
spins (and there are no problems in going to considerably larger systems). 
For small systems the results are in perfect agreement with exact 
diagonalization data.

A ``tempering'' scheme, where $\beta$ is considered as an additional 
discretized dimension of the configuration space \cite{tempering}, was also
implemented in the simulations. Transitions satisfying detailed balance are
carried out between neighboring $\beta$ values. This way, results can be 
obtained on a dense temperature grid with much less effort than by several
fixed-$\beta$ simulations. A temperature spacing $\Delta T/J = 0.01-0.02$ was
used.

\begin{figure}
\includegraphics[width=7cm]{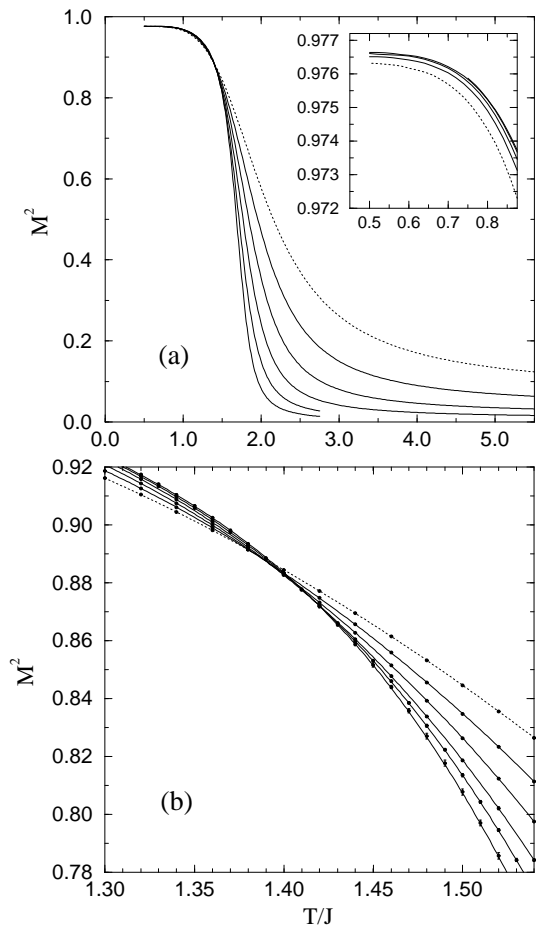}
\caption{(a) Magnetization squared vs temperature for system sizes 
$N=16$ (dotted curve), $32,64$,$128$,$256$ and $512$ (solid curves). The 
statistical errors are smaller then the width of the curves. (b) The same 
quantity on a more detailed scale in the intersection region.
The points with barely visible error bars are the simulation results. The 
curves are  third-order polynomial fits.}
\label{mag}
\end{figure}

Figure \ref{mag}(a) shows results for systems with $N$ up to $512$. At high
temperatures, $M^2$ decreases with increasing $N$, as expected, and there is
a slight increase with $N$ at low $T$. The curves intersect at
$T/J \approx 1.4$. A discontinuous magnetization jump at $T_c$ in the 
thermodynamic limit implies that $M^2$ should become size independent at 
$T_{\rm c}$ for sufficiently large $N$. A notable difference between the 
finite-size behavior of $M(T)$ seen in Fig.~\ref{mag} and the magnetization 
curves for the classical system is that in the latter case the curves do not 
intersect, but the infinite-size value $M(T_c)$ is approached with a 
logarithmic correction \cite{luijten2}. The reason for the different form 
of the finite-size scaling for $h>0$ should be clarified. 

Figure \ref{mag}(b) 
shows in more detail the behavior in the region where the curves intersect. 
The point of intersection moves slowly towards higher $T$ as $N$ increases, 
and larger $N$ would be needed to extract $T_c$ accurately. Based on the data 
presented here $T_{\rm c}/J = 1.42 \pm 0.01$. This can be compared with 
$T_{\rm c} (h=0) \approx 1.53J$ for the classical model \cite{luijten2}. A 
reduction of $T_{\rm c}$ is expected on account of quantum fluctuations for 
$h > 0$. The quite small reduction for $h/J = 0.5$ is consistent with the 
$T \to 0$ magnetization being only slightly reduced from the classical value 
$M(0)=1$. It would clearly also be interesting to study the quantum phase 
transition, but that problem is beyond the scope of this paper.

The high accuracy of these simulations demonstrate that the algorithm indeed 
is very efficient. The computer resources used for this work were quite 
modest; on the order of $200$ CPU hours on an SGI Origin2000. The scaling of 
the CPU time is close to linear in $N$ for the $1/r^2$ interaction, for which 
the interaction sum (\ref{in}) converges rapidly. Only the local updates 
discussed in Sec.~II~B were used in these simulations. The cluster updates 
have been tested as well and improve the performance . The quantum-cluster 
update should be particularly useful for studying the quantum phase 
transition, where there will be a broad distribution of the sizes of the 
clusters constructed in this update.

\section{DISCUSSION}

A new approach to long-range interacting quantum models has here been 
developed within the framework of transverse Ising models. It is important 
to note that the technique can also be generalized to other types of systems,
with the usual caveat of sign problems \cite{ssesign}. What is particular 
about the Ising interaction is that it can be written so that a spin-spin 
term either gives zero or a constant when acting on an arbitrary basis 
state. This is what is needed in order to reduce the interactions to local 
constraints in the SSE formalism. However, the algorithm can easily be 
modified to cases where the diagonal interaction can take several non-zero 
values. The first modification is in the diagonal update. For the Ising model, 
the probability of selecting a given bond (\ref{hd}) is given by a matrix 
element corresponding to the spin pair being in a configuration energetically
favored by the interaction. If the spins are in a non-favored configuration 
(correspinding here to a vanishing matrix element) 
the update is simply rejected. In the general case, the probability to 
use in this update should correspond to the largest diagonal 
matrix element on a given bond, and if the actual configuration 
corresponds to a smaller matrix element the update should be  accepted
only with a probability reflecting this smaller value (i.e., the ratio between
the actual value and the largest matrix element). The quantum-cluster update 
can be modified by using ideas developed within the ``directed-loop'' 
algorithm \cite{sseloop2}. For example, there could be $4$-particle vertex 
processes where the whole vertex is not necessarily reversed as in 
Fig.~\ref{q_cluster}(a). The process could instead 
either go straight through the 
vertex (modifying the vertex only at the entrance and exit legs) or 
``bounce'' back without modifying the vertex at all. The details of how 
this is done in practice will of 
course depend on the types of diagonal and off-diagonal terms in the 
Hamiltonian. The main point to note is that in the SSE approach all the 
information needed to update the vertices is contained in the vertices
themselves, which are always local and can be generated in the diagonal 
update based purely on local decisions.

The transverse Ising simulation algorithm has here been tested on a 
one-dimensional model with long-range interactions decaying as $1/r^2$. The 
program requires almost no modifications for higher-dimensional systems, 
and random interactions are also very easy to implement. Future studies will
have to address how well the method works in practice for a variety of systems
that are more challenging because of frustrated interactions, long-range 
frustrated interactions, or even randomly frustrated long-range interactions. 
For short-range interactions, it would also be interesting to see how the 
SSE quantum-cluster method constructed here compares to the 
transverse Ising cluster method previously developed for continuous-time 
worldline simulations \cite{rieger}.

\acknowledgments{
I would like to thank Patrik Henelius for numerous discussions. This work was 
supported by the Academy of Finland (project 26175).
}

\end{document}